\documentclass[a4paper,11pt]{article}
\usepackage{jinstpub} 
\usepackage{siunitx}
\usepackage{url}
\usepackage{lineno}

\DeclareSIUnit{\ele}{\mbox{$\text{e}^{\text{-}}$}}
\DeclareSIUnit[number-unit-product = ]{\percent}{\%}
\DeclareSIUnit{\raddose}{rad}
\DeclareSIUnit{\tid}{\mbox{\si{\kilo\raddose}}}
\DeclareSIUnit{\niel}{\mbox{\SI{1}{\mega\electronvolt}\ n$_{\mathrm{eq}}$\ \si{\per\cm\squared}}}

\title{First measurements with monolithic active pixel test structures produced in a 65 nm CMOS process}

\author[1]{M. Buckland\note{Now at STFC Daresbury Laboratory, Keckwick Ln, Warrington WA4 4AD, UK.}\collaboration[c]{on behalf of the ALICE collaboration}}
\affiliation{University \& INFN Trieste,\\Via A. Valerio 2, 34127 Trieste, Italy}

\emailAdd{matthew.buckland@stfc.ac.uk}

\abstract{The Inner Tracking System (ITS) of the ALICE experiment at CERN will undergo an upgrade during the LHC long shutdown 3, in which the three innermost tracking layers will be replaced. This upgrade, named the Inner Tracking System 3 (ITS3), employs stitched wafer-scale Monolithic Active Pixel Sensors fabricated in a \qty{65}{\nm} CMOS process. The sensors are \qty{260}{\mm} in length and thinned to less than \qty{50}{\um} then bent to form truly half-cylindrical half-barrels. The feasibility of this process for the ITS3 was explored with the first test production run (MLR1) in 2021, whose goal was to evaluate the charged particle detection efficiency and the sensor performance under non-ionising and ionising radiation up to the expected levels for ALICE ITS3 of \qty{e13}{\niel} (NIEL) and \qty{10}{\kilo\gray} (TID). Three sensor flavours were produced to investigate this process: Analog Pixel Test Structure (APTS), Circuit Exploratoire 65 (CE65) and Digital Pixel Test Structure (DPTS).

This contribution gives an overview of the MLR1 submission and test results, describing the different sensor flavours and presenting the results of the performance measurements done with particle beams for various chip variants and irradiation levels.}


\begin{document}
\maketitle
\flushbottom

\section{Introduction}
\label{sec:intro}

The importance of Monolithic Active Pixel Sensors (MAPS) for use in high energy physics experiment vertex and tracking detectors has been established in the last decade. The most recent implementation of a large-area MAPS detector was for the ALICE Inner Tracking System 2 (ITS2)~\cite{ALICEITS2} at CERN, which used the ALPIDE chip~\cite{ALPIDE-proceedings-2,ALPIDE-proceedings-1}. This chip was fabricated in the TowerJazz \qty{180}{\nm} CMOS process and showed excellent performance in terms of detection efficiency (\qty{\gg 99}{\percent}) and spatial resolution (about \qty{5}{\um}).

During the LHC Long Shutdown 3 (2026-2028), the ITS2 will undergo an upgrade called the ITS3, where the three innermost tracking layers will be replaced. The ITS3 employs wafer-scale MAPS with a length of \qty{260}{\mm} that are thinned to \qty{<50}{\um} and bent to radii of \qty{18}{\mm}, \qty{24}{\mm}, and \qty{30}{\mm} to form cylindrical half-barrels. To obtain sensors of this length, a process called stitching is employed. In this process the reticles of the CMOS imaging process are joined together to produce a larger single sensor, removing the need for electrical services in the detector. By utilising the natural stiffness of the cylindrical geometry, the majority of the mechanical support can also be removed, using only carbon foam spacers as support structures. Finally, the very low power consumption (\qty{<20}{\milli\watt\per\cm^2}) of the chip will allow the detector to be cooled by air. These reductions in the material within the sensitive volume will enable the ITS3 to have a very low material budget of \qty{<0.05}{\percent} X$_{0}$ per layer. Altogether, these upgrades will provide exceptional tracking and vertexing capabilities leading to an improvement in the pointing resolution of the current detector by a factor of two.

Due to the challenging design that the ITS3 poses, the Tower Partners Semiconductor Co.~(TPSCo) \SI{65}{\nm} CMOS imaging process~\cite{tower} was chosen as the starting point. The first submission in the \qty{65}{\nm} CMOS process, in conjunction with the CERN~EP~R\&D on monolithic pixel sensors~\cite{eprnd}, was called MLR1 and contains many different test structures to fully explore the CMOS process.

\section{MAPS in the \qty{65}{\nm} CMOS imaging process}
\label{sec:mlr1}

By using a smaller CMOS process node compared to the \SI{180}{\nm} CMOS process of the ITS2, more possibilities will be open, such as:

\begin{itemize}
    \item the ability to have complete coverage along the beam axis (\textit{z}-direction) in the detector with a single stitched sensor produced on \qty{300}{\mm} wafers;
    \item lower power consumption when moving to deeper submicron processes.
\end{itemize}

However, moving to a new process also provides challenges, such as optimising the design for sensor yield, the charge collection, and testing and verifying the radiation harness.

\begin{figure}[htbp]
    \centering
    \includegraphics[width=.55\textwidth]{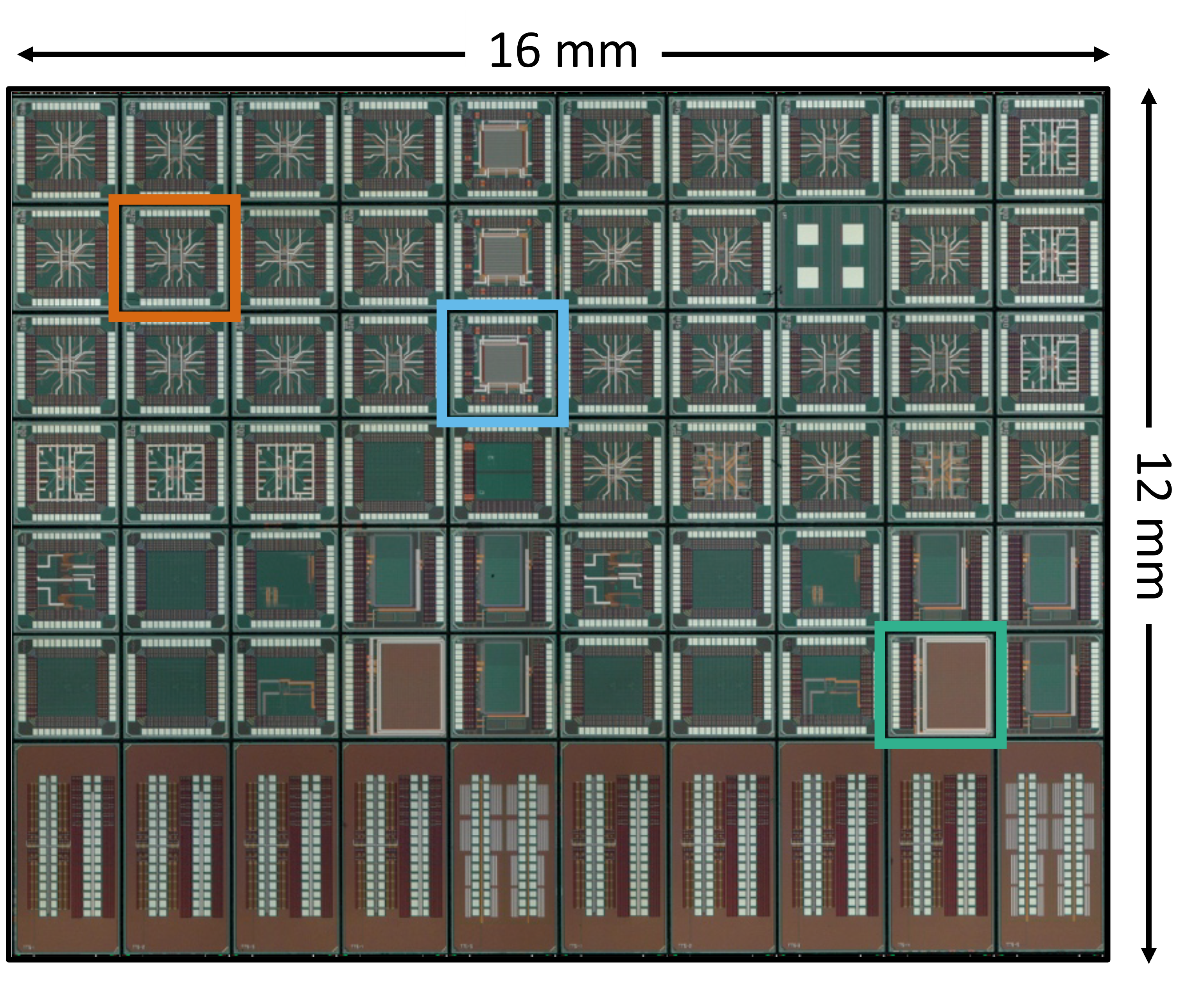}
    \caption{The MLR1 reticle floor plan highlighting the APTS (orange), CE65 (green) and DPTS (blue) chips.}
    \label{fig:mlr1}
\end{figure}

Three main sensor flavours produced are the Analog Pixel Test Structure (APTS), Circuit Exploratoire 65 (CE65), and Digital Pixel Test Structure (DPTS), each measuring \qtyproduct{1.5x1.5}{\mm} in size and highlighted in Fig.~\ref{fig:mlr1}.

\begin{figure}[htbp]
    \centering
    \includegraphics[width=.98\textwidth]{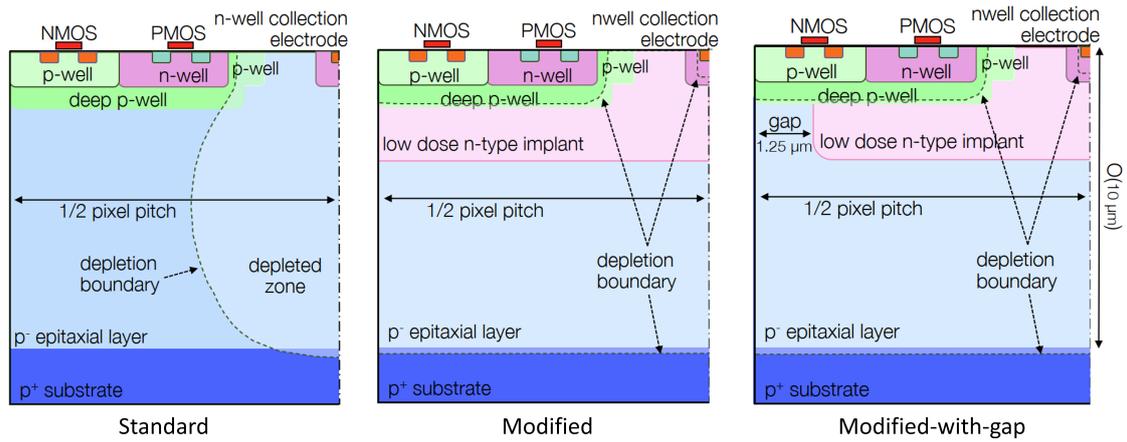}
    \caption{The three process options implemented in the MLR1 chips: standard (left), modified (middle), and modified-with-gap (right).}
    \label{fig:x-secs}
\end{figure}

In addition to three sensor flavours, three process options were also explored to investigate the charge collection properties of the CMOS process. These processes, shown in Fig.~\ref{fig:x-secs}, are called standard, modified, and modified-with-gap and are similar to those used in the \SI{180}{\nm} CMOS process~\cite{modproc}. In the standard process, the epitaxial layer is only partially depleted, so some of the charges will undergo diffusion, and this process is expected to have the largest charge sharing. For the modified process, a low-dose n-type implant is added across the length of the pixel. This enables the epitaxial layer to be fully depleted and increases the lateral electric field to the collection diode. Thus, it better collects the signal charge and accelerates the carriers towards the collection diode. This lateral electric field is further increased in the modified-with-gap process where a gap in the low-dose n-implant is added at the edges of the pixel and results in a similar depletion region to the modified process.

The main goals of the MLR1 submission were to verify that the detection efficiency was \qty{99}{\percent} and that the radiation hardness could reach the expected levels for the ITS3 (\qty{e13}{\niel} and \qty{10}{\kilo\gray}).

\section{Analog Pixel Test Structure}
\label{sec:apts}

The APTS incorporates a \numproduct{6x6} pixel matrix with direct analogue readout on the central \numproduct{4x4} pixels. Two versions of the output buffer were implemented: a source-follower (APTS-SF) and a fast operational amplifier (APTS-OA) whose focus was on time resolution. In addition, the sensor was produced in four different pixel pitches ranging from \qty{10}{\um} to \qty{25}{\um}. The goal of the APTS was to explore the different sensor designs and processes.

\begin{figure}[htbp]
    \centering
    \includegraphics[width=.99\textwidth]{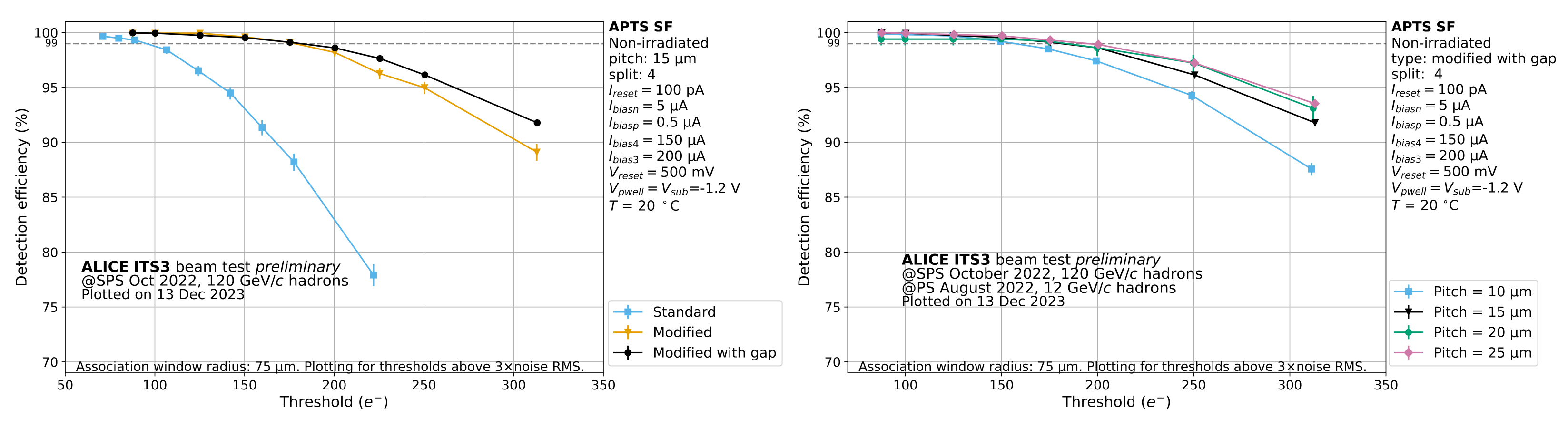}
    \caption{Comparison of the detection efficiency vs. threshold for an APTS-SF sensor for the three processes (left, see text for details) and different pixel pitches (right).}
    \label{fig:apts-eff} 
\end{figure}

\begin{figure}[htbp]
    \centering
    \includegraphics[width=.55\textwidth]{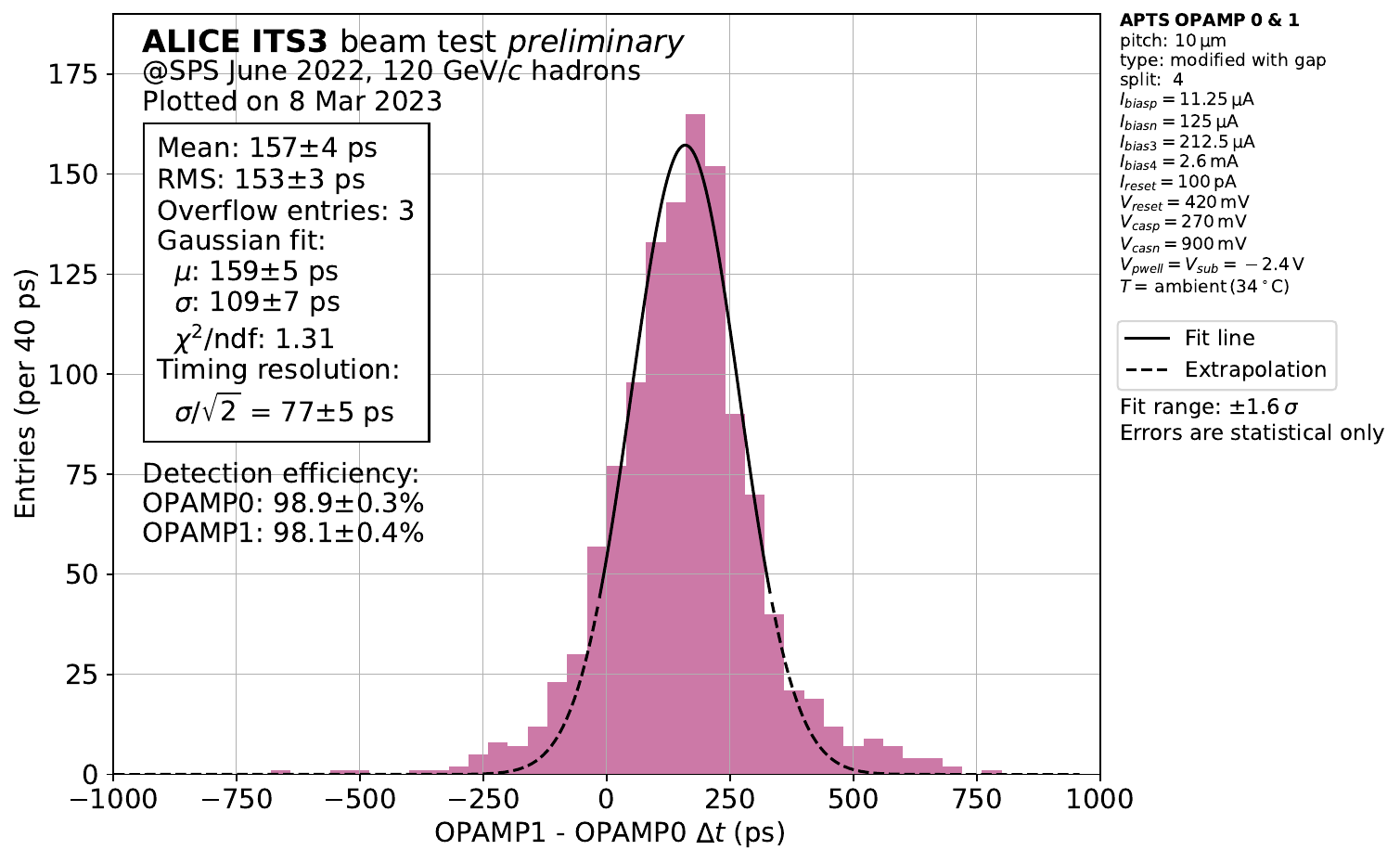}
    \caption{The time residual distribution of two APTS-OA sensors fitted with a Gaussian function to extract the timing resolution.}
    \label{fig:apts-oa} 
\end{figure}

The in-beam measurements of the APTS-SF shown in Fig.~\ref{fig:apts-eff} (left) demonstrate the impact of the different process types on the detection efficiency. While all three can reach the desired \qty{99}{\percent}, it is clear that the standard process has a reduced performance at larger thresholds compared to the other two due to the improved charge collection in the modified processes. The modified-with-gap process shows the largest detection efficiency over the whole measured threshold range. Comparing the detection efficiency for different pitches, Fig.~\ref{fig:apts-eff} (right), it can be seen that below a threshold of \qty{200}{\ele}, there is minimal difference among the pitches. However, above this value, larger pitches result in higher efficiencies. Timing measurements were also performed with in-beam measurements of two APTS-OA, resulting in a timing resolution of ($77\pm5$)~\si{\pico\s}, Fig.~\ref{fig:apts-oa}.

\section{Circuit Exploratoire 65}
\label{sec:ce65}

The CE65 is a ``large"-area chip consisting of an analogue rolling shutter readout with an integration time of \qty{50}{\us}. One type of the pixel matrix consists of \numproduct{64x32} pixels implemented with a pixel pitch of \qty{15}{\um} and split into three subvariants that differ by their in-pixel amplifier: AC, DC or SF. The other type of pixel matrix contains \numproduct{48x32} pixels implemented with a pitch of \qty{25}{\um}. The goal of the CE65 was to study the pixel matrix uniformity.

\begin{figure}[htbp]
    \centering
    \includegraphics[width=.65\textwidth]{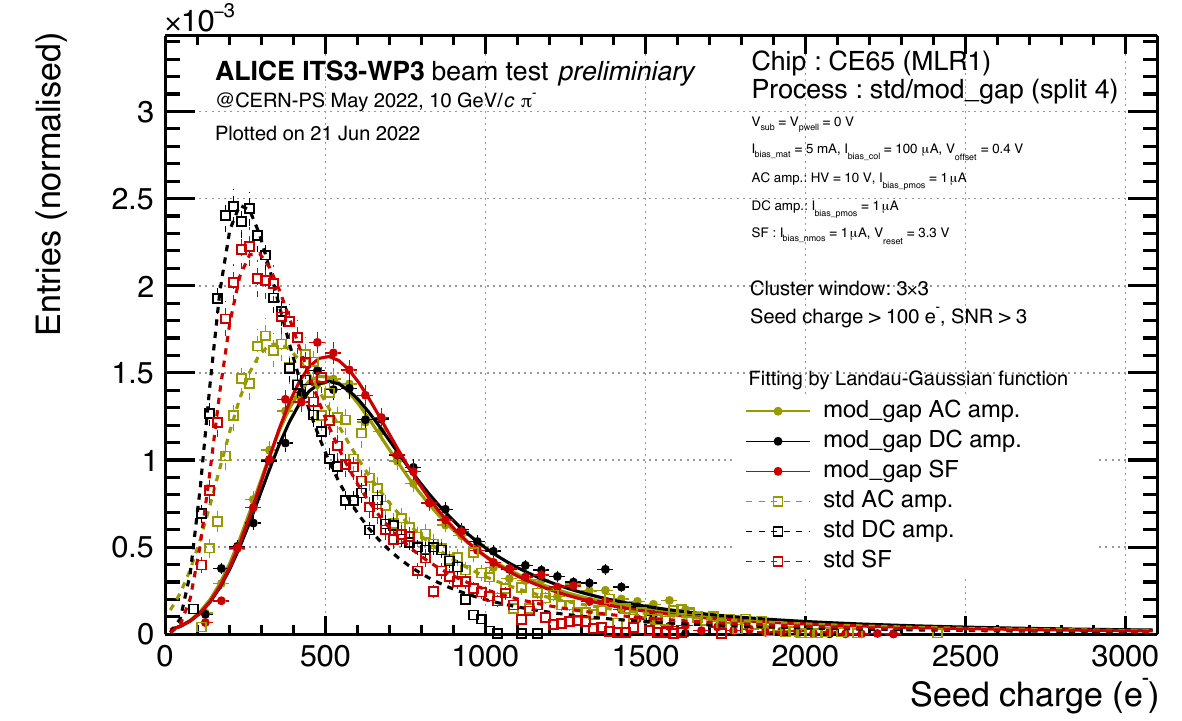}
    \caption{The in-beam seed pixel distribution comparing the response of the different CE65 variants.}
    \label{fig:ce65} 
\end{figure}

Figure \ref{fig:ce65} shows the seed pixel distributions obtained from in-beam measurements of the different CE65 variants. There is a clear distinction between the standard and the modified-with-gap processes, with the latter having a larger most probable value (MPV), signifying a larger charge collection depth. The difference between the amplifiers is minimal for the modified-with-gap process. However, for the standard process, it can be seen that the AC-coupled amplifier has a larger MPV.

\section{Digital Pixel Test Structure}
\label{sec:dpts}

The DPTS features a \numproduct{32x32} pixel matrix with a pitch of \qty{15}{\um} implemented in the modified-with-gap process and contains a full digital front-end with asynchronous readout \cite{AGLIERIRINELLA2023168589}. The sensor is controlled by a set of external reference currents and voltages and read out via a current mode logic (CML) output \cite{cecconi-twepp,10196055}. All the pixels are read out simultaneously via a differential digital output that time encodes the pixel position and Time-over-Threshold (ToT). The goal of the DPTS was to study the in-pixel full-digital front end.

\begin{figure}[htbp]
    \centering
    \includegraphics[width=.75\textwidth]{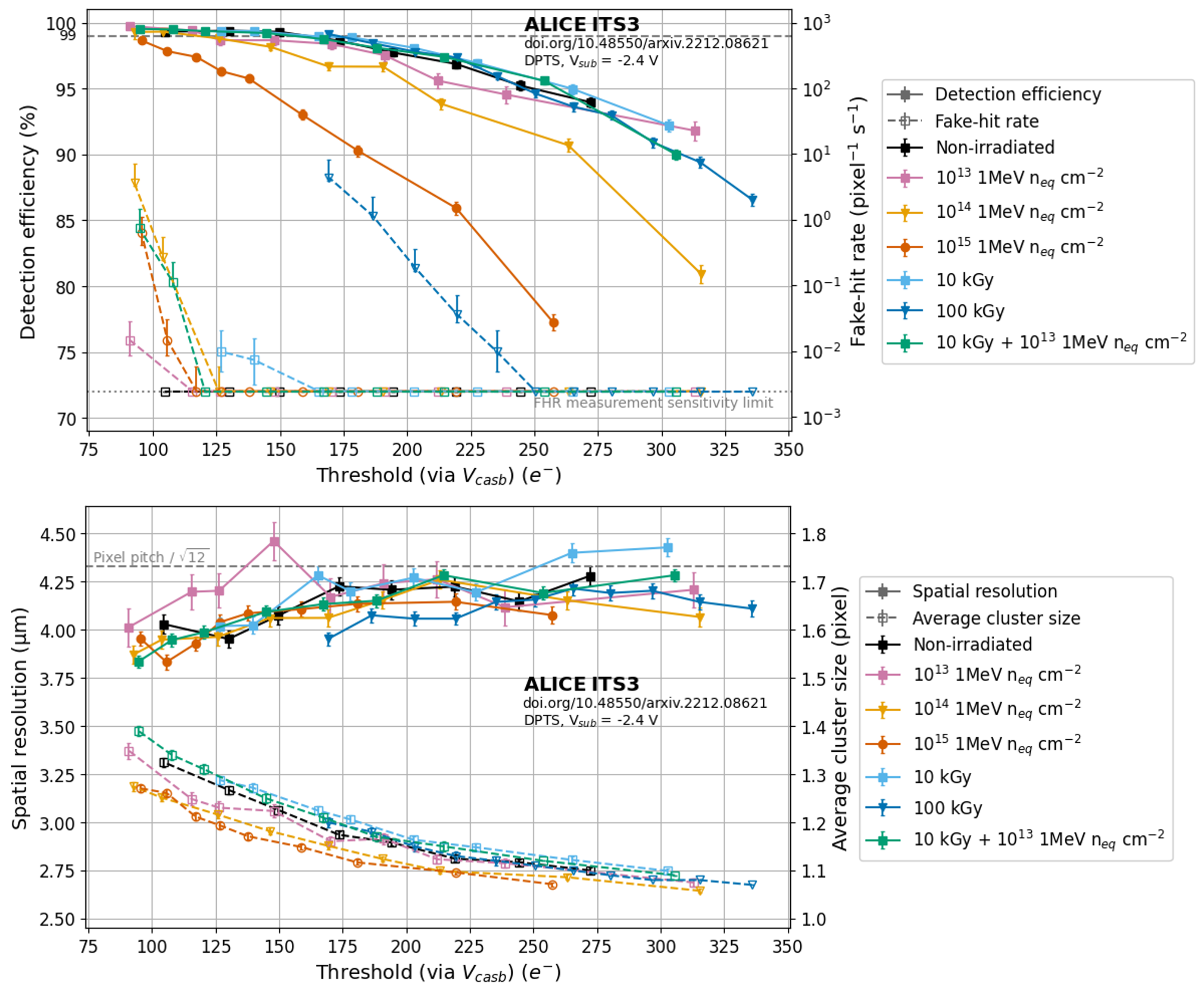}
    \caption{Top: the detection efficiency (filled symbols, solid lines, left axis) and fake-hit rate (open symbols, dashed lines, right axis) vs. threshold. Bottom: the spatial resolution (filled symbols, solid lines, left axis) and average cluster size (open symbols, dashed lines, right axis) vs. threshold. Both plots are for DPTS chips irradiated to various levels \cite{AGLIERIRINELLA2023168589}.}
    \label{fig:dpts}
\end{figure}

The performance of DPTS chips for various irradiation levels taken at a temperature of +\SI{20}{\celsius} was evaluated using in-beam measurements of positive hadrons at \qty{10}{\giga\electronvolt\per c}, the results of which are shown in Fig.~\ref{fig:dpts}. For all irradiation levels, the sensor shows an excellent detection efficiency of \qty{99}{\percent} and a spatial resolution below the binary resolution (pixel pitch / $\sqrt{12}$) while preserving a fake-hit rate below \num{10}~pixel$^{-1}$~s$^{-1}$. For the detection efficiency, it can be seen that non-ionising irradiation leads to a decrease in detection efficiency while ionising irradiation leads to an increase in the fake-hit rate. For the spatial resolution, there is a negligible impact of the irradiation on the performance of the tested irradiation levels. Whereas the average cluster size shows a slight decrease with increasing non-ionising dose.

\begin{figure}[htbp]
    \centering
    \includegraphics[width=.85\textwidth]{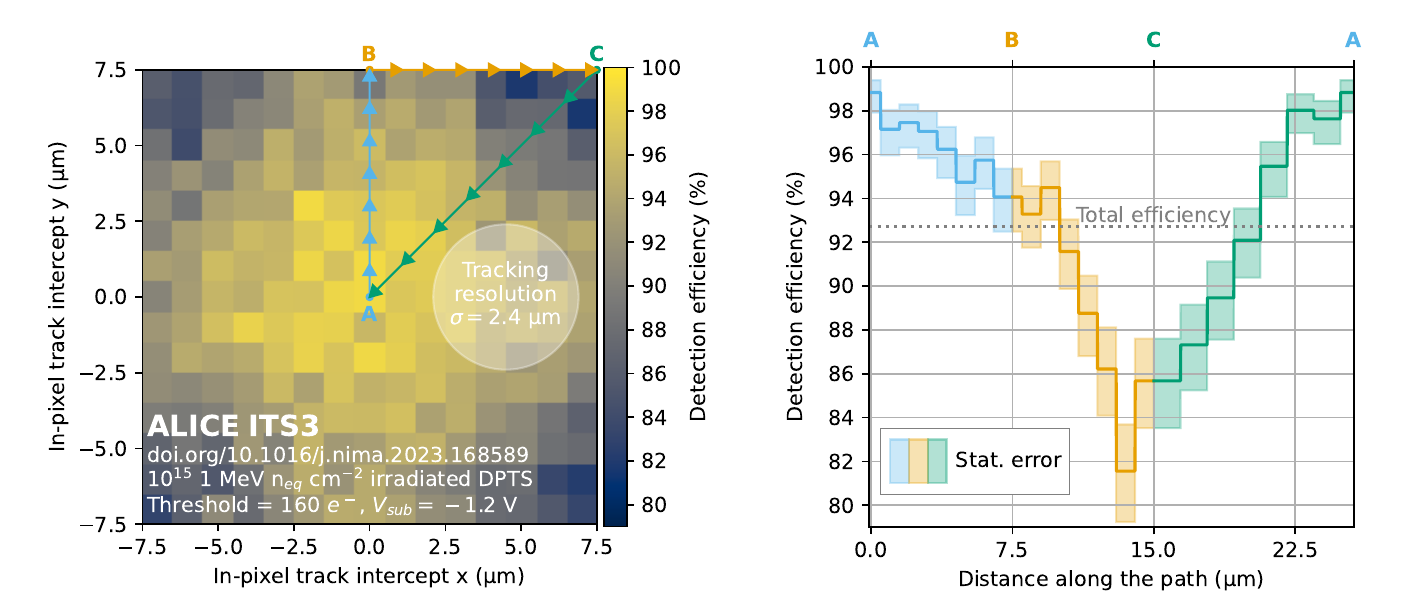}
    \caption{The in-pixel detection efficiency for a DPTS sensor irradiated to \qty{e15}{\niel} with threshold set to \qty{160}{\ele}, measured with \qty{10}{\giga\electronvolt\per c} positive hadrons. The grey circle represents the tracking resolution of the telescope used to reconstruct the tracks~\cite{AGLIERIRINELLA2023168589}.}
    \label{fig:dpts-in-pix} 
\end{figure}

In-beam measurements using positive hadrons at \qty{10}{\giga\electronvolt\per c} were also used to investigate the cause of the detection efficiency loss in the sensor by measuring the particle hit position within a pixel. The detection efficiency was studied as a function of the reconstructed track position relative to the nearest pixel centre for a sensor irradiated to \qty{e15}{\niel} with a threshold of \qty{160}{\ele}, as shown in Fig.~\ref{fig:dpts-in-pix}. It can be seen that the further the track is away from the collection diode, in the centre of the pixel, the smaller the detection efficiency is, an effect becoming particularly acute in the corners of the pixel.

\section{Summary}
\label{sec:sum}

The performance of the MLR1 chips was evaluated through extensive characterisation in the laboratory and with in-beam measurement. The measurements show that the MLR1 was a success thanks to the large number of operational prototypes that allow the parameter space of the CMOS process to be mapped out. Furthermore, the MLR1 structures exhibit excellent performance in terms of detection efficiency (\qty{>99}{\percent}) and spatial resolution (3-4.5~\si{\um}) from the in-beam measurements for all three sensor flavours considered. The radiation hardness is demonstrated by the sensors maintaining a detection efficiency of \qty{99}{\percent} for chips irradiated with a dose at the expected ITS3 levels, \qty{e13}{\niel} (NIEL) and \qty{10}{\kilo\gray} (TID). The radiation hardness actually exceeds this goal as the desired performance is maintained even for sensors irradiated up to \qty{e15}{\niel} and operated at +\SI{20}{\celsius}. In addition, the APTS-OA has demonstrated a time resolution of ($77\pm5$)~\si{\pico\s}.

From the results of the MLR1, the detection efficiency and radiation hardness have been validated and represent an important milestone in the R\&D for ALICE ITS3. The next step towards a wafer-scale bent sensor is the second submission in the \qty{65}{\nm} process designated ER1, whose goal is the validation of stitching and yield via the full-scale sensor prototypes.

\bibliographystyle{JHEP}
\bibliography{biblio.bib}

\end{document}